\documentclass{llncs}

\usepackage{amsfonts}
\usepackage{amsmath}
\usepackage{amssymb}

\title{Fingerprinting Logic Programs}

\author{Alexander Serebrenik\inst{1} \and Wim Vanhoof\inst{2}}

\institute{
Technische Universiteit Eindhoven\\
Den Dolech 2,
P.O. Box 513 \\
5600 MB  Eindhoven,
The Netherlands\\
\email{a.serebrenik@tue.nl}\\
\ \\
\and
University of Namur\\
Rue Grandgagne, 21\\
B-5000 Namur, Belgium\\
\email{wva@info.fundp.ac.be}
}

\newcommand{\alexander}[1]{}
\newcommand{\wim}[1]{}
\newcommand{\multiset}[1]{\ensuremath{\{#1\}}}

\newcommand{\vars}{\ensuremath{{vars}}}
\newcommand{\head}{{\ensuremath{{head}}}}
\newcommand{\body}{{\ensuremath{body}}}
\newcommand{\pred}{{\ensuremath{pred}}}
\newcommand{\dom}{{\ensuremath{dom}}}
\newcommand{\Clauses}{{\ensuremath{Clauses}}}

\newcommand{\gp}{{\ensuremath{GP}}}

\newcommand{\gpA}{{\ensuremath{GP_{\!\!\mathcal A}}}}
\newcommand{\cpA}{{\ensuremath{CP_{\!\!\mathcal A}}}}
\newcommand{\ppA}{{\ensuremath{PP_{\!\!\mathcal A}}}}

\newenvironment{SProg}
     {\begin{small}\begin{tt}\begin{tabular}[c]{l}}%
     {\end{tabular}\end{tt}\end{small}}

\newcommand{\qin}{\hspace*{0.15in}}

\begin{document}

\maketitle

\begin{abstract}
In this work we present work in progress on functionality duplication detection in logic programs. Eliminating duplicated functionality recently became prominent in context of refactoring. We describe a quantitative approach that allows to measure the ``similarity'' between two predicate definitions. Moreover, we show how to compute a so-called ``fingerprint'' for every predicate. Fingerprints capture those characteristics of the predicate that are significant when searching for duplicated functionality. Since reasoning on fingerprints is much easier than reasoning on predicate definitions, comparing the fingerprints is a promising direction in automated code duplication in logic programs.
\end{abstract}

\section{Introduction}
Refactoring~\cite{fowler:book} is a  source-to-source program transformation that
changes program structure and organization, but not program functionality.
The major aim of refactoring is to improve readability, maintainability
and extensibility of the existing software. Refactoring has been shown to be
profitable both for developing new software and for maintaining existing 
software.
Refactoring~\cite{Mens:Tourwe} consists of series of small 
transformation steps, also
known as refactorings. For each step, an appropriate code fragment and an 
appropriate transformation have to be chosen, the transformation has to be 
executed and evaluated. In this paper we restrict our attention to the first 
step, namely identifying potential for transformation 
application. A number of refactorings aim at eliminating duplicated 
code (or better: duplicated functionality) and therefore automatic detection of code duplication 
becomes a necessity. 

Code duplication can be caused by a number of reasons. First of all,
it can result from unfamiliarity of the developer with the existing
code body. Second, the ``copy and paste'' technique is commonly used when 
the existing functionality has to be slightly adapted. Although in this case 
one usually does not end up literally duplicating the code, the changes 
introduced by adaptation are usually relatively minor and a generalization 
of the original and the adapted fragments can be often proposed. Finally,
code duplication might result from a polyvariant program 
analysis~\cite{DBLP:conf/lopstr/OchoaPH05}.     

This being true for any programming paradigm we 
concentrate on logic programming (LP). 
While code duplication detection for imperative and object-oriented 
programming languages has been often studied in the past
~\cite{Baker95b,DBLP:conf/icsm/BaxterYMSB98,DBLP:conf/icsm/DucasseRD99,%
DBLP:journals/tse/KamiyaKI02,DBLP:conf/sas/KomondoorH01,%
DBLP:journals/ase/KontogiannisDMGB96,DBLP:conf/icsm/MayrandLM96,%
DBLP:conf/wcre/RiegerDL04}, this topic has attracted less research attention 
in the logic programming community. To the best of our knowledge the 
only results on code duplication in LP are due to Vanhoof~\cite{wim:lopstr2004}
motivated by the study of refactoring techniques for logic 
programs~\cite{alex:refactoring}. 
 
In a logic programming setting, we say that two predicates are duplicates if their definitions are identical up to a consistent renaming of variables and a permutation of the arguments.
\alexander{...if their definitions are identical up to a consistent renaming of variables, function and predicate symbols. Wat denk je ervan? Als je accord gaat moet ook de bespreking na het voorbeeld aangepast worden. Het is natuurlijk zwakker dan ``the same relation'' maar om over de computed answers te redeneren moeten we nog wat doen.}
\wim{'k snap waar je naartoe wil, maar da's 't ook niet helemaal want enerzijds hernoemen we geen functoren en 't is alleen het gedefinieerde predikaat dat hernoemd mag worden maar anderzijds doen we wel meer dan renamen. Wat denk je van bovenstaande zin? Die bevat voldoende informatie voor 't voorbeeld, en wordt nadien uitgebreid. }
Consider for example the predicates {\tt append/3} and {\tt concat/3} depicted below:
\[\mbox{
\begin{SProg}
append([],L,L).\\
append([X|Xs],Y,[X|Zs]):- append(Xs,Y,Zs).\\
 \\
concat(L,[],L).\\
concat([E|Zs],[E|Es],Y):- concat(Zs,Es,Y).\\
\end{SProg}}\]
Even though the above predicate definitions are not literal copies of one another, intuitively it is clear that they are meant to perform the same operation, i.e. list concatenation.
\wim{wat denk je?}
In addition to support variable renaming and argument permutation, we would like to have our notion of code duplication to be to some extent independent from the order of the clauses and the order of the atoms (including the unifications) in the clause bodies. 
As a rather trivial example, reconsider the definition of the {\tt append/3} predicate, this time written in a kind of normal form, where the unifications have been moved from the head to the body:
\[\mbox{
\begin{SProg}
append(X,Y,Z):- X = [], Z = Y.\\
append(X,Y,Z):- X = [Xe|Xs], Z = [Xe|Zs], append(Xs,Y,Zs).\\
\end{SProg}}\]
In the definition above, one could easily switch the order of the unifications while the resulting predicate could still be considered a duplicate of the original {\tt append/3}. This degree of liberty one has in organizing the code makes approaches based on textual pattern-matching like \cite{Baker95b,DBLP:conf/icsm/DucasseRD99} less suited for our purposes. Moreover, unlike imperative programming languages with a well-developed set of control keywords (if, while, repeat, switch, etc.) control structures in Prolog are less explicit. This hinders the application of  textual pattern-matching approaches to logic programming.

As a second example, let us consider two predicates that are not duplicates but that are nevertheless {\em similar} in the sense that they contain some common functionality:
\[\mbox{\begin{SProg}
rev\_all([],[]).\\
rev\_all([X|Xs],[Y|Ys]):- reverse(X,Y), rev\_all(Xs,Ys).\\
\\
add1\_and\_sqr([],[]).\\
add1\_and\_sqr([X|Xs],[Y|Ys]):- N is X + 1, Y is N*N, add1\_and\_sqr(Xs,Ys).
\end{SProg}}\]
These definitions implement two different relations: {\tt rev\_all} reverses all the elements of an input list, while {\tt add1\_and\_sqr} transforms each element $x$ of an input list into $(x+1)^2$. They nevertheless have a common core and if we assume a language with higher-order capabilities (as for example in \cite{warren:ho}), one can extract or {\em generalize} the common functionality into a {\tt map/3}-like predicate and translate each call to {\tt rev\_all/2} and {\tt add1\_and\_sqr/2} into an appropriate call to {\tt map/3}, providing the code specific to {\tt rev\_all/2} or {\tt add1\_and\_sqr/2} as an argument.

In \cite{wim:lopstr2004} we have given a formal characterization of code duplication in the sense outlined above. While the associated analysis, which basically tries to establish an isomorphism between two predicate definitions by comparing every possible pair of subgoals, {\em can} be used to search for duplication, its complexity renders it hard if not impossible to use in practice. Worse, the analysis is not quantitative: even though it may find some common functionality between two or more predicate definitions, it has no way of indicating {\em how similar} the definitions are. Yet, this is important if the analysis is to be used in a practical tool since not every pair of predicates that share some common functionality is susceptible to generalization.

In this work, we revise the notion of code duplication for logic programs.
In a first step, we formally define a {\em quantitative} measure that reflects the {\em similarity} between two predicate definitions. In contrast with earlier work \cite{wim:lopstr2004}, this allows us not only to detect predicate definitions that are duplicates of one another, but it also provides us with a meaningful indication about {\em how much} code is common between two predicate definitions. In a second step, we show how to compute a so-called {\em fingerprint} for every predicate in the software system under consideration. Such a fingerprint captures in a single value those characteristics of a predicate that are significant when searching for duplicated (or common) functionality while it abstracts those characteristics that are much less relevant during the search. Our domain of fingerprint values is such that 1) duplicated predicates are mapped onto the same fingerprint, and 2) an order relation can be defined on fingerprint values that reflect the degree of ``similarity'' between the corresponding predicates. Predicates whose fingerprints are ``close'' to one another in the order are likely to share a common structure and hence are potential candidates for generalization. 


\section{Basic Definitions}

In what follows, we assume the reader to be familiar with the basic logic programming concepts as they are found, for example, in \cite{apt:lpth,lloyd:found}. As usual, variable names will be represented by uppercase symbols $X,Y,\ldots$ whereas predicate and function symbols by lowercase letters. Unless noted otherwise, we will use $p,q,r,\ldots$ to refer to predicate names and $f,g,h,\ldots$ to refer to function names.

We restrict ourselves to definite programs. In particular, we consider a program to be defined as a set of clauses of the form $H\leftarrow B_1\wedge \ldots \wedge B_n$ with $H$ an atom and $B_1\wedge \ldots \wedge B_n$ a conjunction of atoms. In our examples we also use the Prolog-style notation for clauses, i.e., we write $\tt{:-}$ instead of $\leftarrow$ and $\tt{,}$ instead of $\wedge$. A {\em goal} is a conjunction of atoms. Given a goal $B_1\wedge \ldots\wedge B_n$, we write $\multiset{B_1,\ldots,B_n}$ to represent the {\em multiset} of atoms occurring in it. A goal $A_1\wedge \ldots\wedge A_m$ is called a {\em subgoal} of a goal $B_1\wedge \ldots\wedge B_n$ if the multiset
 $\multiset{A_1,\ldots,A_m}$ is a submultiset of the multiset $\multiset{B_1,\ldots,B_n}$. 
Given a particular clause $c$, we denote by $\head(c)$ and $\body(c)$ the atom and the conjunction of atoms that constitute the head and the body of the clause, respectively. For an atom $A$, we denote by $\pred(A)$ the predicate symbol used in $A$. 
Given a predicate symbol $p/n$ we denote by $\Clauses(p/n)$ the set of clauses $c$ such that $\pred(\head(c))$ coincides with $p/n$. 

Predicates can be mutually recursive. Therefore, rather than considering individual predicate definitions, we will consider  strongly connected components in the predicate dependency graph. Since strongly connected components can be seen as equivalence classes with respect to the ``depends on'' relation~\cite{aptbook} the strongly connected component (SCC) of a predicate $p/n$ is denoted $[p/n]$. Given a strongly connected component $[p/n]$, we denote by 
$\Clauses([p/n])$ the union of the sets of clauses $\Clauses(q/m)$ for all 
$q/m\in [p/n]$. 
In what follows we will often drop the arity from predicate symbols and we will write $p$ (and likewise $[p]$) instead of $p/n$ (and $[p/n]$).

We will often need to refer to those atoms in the body of a clause 
that represent a (direct or indirect) recursive call. 
Given a clause $c$ an atom $B_i$ in $\body(c)$ is called a
{\em recursive call} if $\pred(B_i)$ belongs to $[\pred(\head(c))]$.
Moreover, we will represent a clause $c\in\Clauses([p])$ as
\[A_0\leftarrow Q_1\wedge A_1\wedge \ldots\wedge Q_k\wedge A_k\wedge Q_{k+1}\]
where $A_i$ ($0\le i\le k$) is a recursive call and $Q_i$ ($1\le i\le k+1$) is a (possibly empty) conjunction of atoms such that none of the conjuncts is a recursive call. 

A {\em variable renaming} is a bijective mapping from variables onto variables. For any mapping $f:X\mapsto Y$, we denote by $f_{|_{D}}$ the restriction of the mapping to the domain $D\subseteq X$. The inverse of any mapping $f$ is denoted by $f^{-1}$. We use the notation $\{x_1/y_1,\ldots,x_n/y_n\}$ to explicitly represent a mapping $f:X\mapsto Y$ with $\dom(f)=\{x_1,\ldots,x_n\}$ and $y_i = f(x_i)$ $\forall i$.

For any syntactic entity $E$ (be it a term, atom, goal or clause), we use $\vars(E)$ to denote the set of variables occurring in $E$. As usual, a {\em substitution} is defined as a finite mapping from distinct variables to terms. Substitutions are usually denoted by Greek letters such as $\sigma, \theta, \ldots$ and for a syntactic entity $E$ and substitution $\theta$ we denote by $E\theta$ the result of applying $\theta$ to $E$. Given two syntactic entities $E_1$ and $E_2$ a {\em generalization} of $E_1$ and $E_2$ is a syntactic entity $E$ such that there exist substitutions $\sigma_1$ and $\sigma_2$ where $E\sigma_1 = E_1$ and $E\sigma_2 = E_2$. A {\em most specific generalization (msg)} of $E_1$ and $E_2$ is a generalization $E$ of $E_1$ and $E_2$ such that for any generalization $E'$ of $E_1$ and $E_2$ there exists a substitution
$\sigma$ such that $E = E'\sigma$. One can show the existence of a {\em unique} most specific generalization (up to variable renaming).

\newcommand{\nodes}{nodes}
\newcommand{\term}{\ensuremath{\mbox{\em Term}}}
We characterize the {\em size} of a term by counting the number of (internal) 
nodes in the term's tree representation and the number of leaves, corresponding to constants. To that extent, if we denote by 
\term\ the set of terms, we define the following mapping 
$\nodes:\term\mapsto\mathbb{N}$:
\[\begin{array}{lll}
\nodes(X)  & = & 0\\
\nodes(f(t_1,\ldots,t_n)) & = & 1 + \sum_{i=1}^n \nodes(t_i).
\end{array}\]
Since goals and clauses can be considered terms constructed by the $\wedge$ and $\leftarrow$ functors, we will use the above measure to characterize the size of goals and clauses as well. By extension, when considering a strongly connected component $[p]$, we define $\nodes([p]) = \sum_{c\in\Clauses([p])}\nodes(c)$.


\section{Identifying duplication and similarity}

\subsection{Comparing goals}

In what follows, we define a quantitative measure that represents the {\em degree of similarity} between two goals. As a starting point, let us define a measure that compares two goals in a purely syntactic way, by simply counting the number of common nodes in the goals' term representations.

\begin{definition}
Given a pair of goals $Q_1 = A_1\wedge \ldots\wedge A_n$ and $Q_2 = A'_1\wedge \ldots\wedge A_n'$, we define the {\em strict commonality} between $Q_1$ and $Q_2$ as the natural number, denoted $c(Q_1,Q_2)$, which is defined as follows:
\[\begin{array}{lll}
c((A_1\wedge A_2\wedge \ldots\wedge A_n), (A'_1\wedge A'_2\wedge \ldots\wedge A'_n)) & = & 1 + c(A_1, A'_1)\\
& &  + c(A_2\ldots\wedge A_n, A'_2\ldots\wedge A'_n)\\
c(p(t_1,\ldots,t_k),p(t'_1,\ldots,t'_k)) & = & 1 + \sum_{i=1}^k c(t_i,t'_i)\\
c(p(t_1,\ldots,t_k),q(s_1,\ldots,s_l)) & = & 0\\
c(f(t_1,\ldots,t_k),f(t'_1,\ldots,t'_k)) & = & 1 + \sum_{i=1}^k c(t_i,t'_i)\\
c(f(t_1,\ldots,t_k),g(s_1,\ldots,s_l)) & = & 0\\
c(f(t_1,\ldots,t_k),X) & = & 0\\
c(X,X) & = & 1\\
c(X,Y) & = & 0\\
c(X,g(s_1,\ldots,s_l)) & = & 0
\end{array}\]
\end{definition}

\begin{example}
\label{ex:similarity:goals}
Consider the following goals: $Q_1=\mbox{\tt p(f(X),g(Y,h(Z,a))), q(Z,X)}$ 
and $Q_2=\mbox{\tt p(f(T),g(T,h(Z,b))), q(Z,T)}$. Then, the strict commonality between them is 
$c(Q_1,Q_2) = 8$.
\end{example}

Note that for two goals $Q_1$ and $Q_2$, the strict commonality $c(Q_1,Q_2)$ quantifies the amount of structure that would be preserved upon taking the most specific generalization of $Q_1$ and $Q_2$. 

\begin{lemma}
Let $Q_1$ and $Q_2$ be as required in Definition~\ref{ex:similarity:goals}.
Then, $c(Q_1,Q_2) = nodes(msg(Q_1,Q_2)) + \delta$ where $\delta$ represents the number of occurrences of identical variables that appear in identical positions in the tree representation of $Q_1$ and $Q_2$.
\end{lemma}
\begin{proof}
The proof is done inductively on the structure of $Q_1$ and $Q_2$. 
\end{proof}

\begin{example}
For goals $Q_1$ and $Q_2$ in Example~\ref{ex:similarity:goals} we have
\[msg(Q_1,Q_2) = \mbox{\tt p(f(\_),g(\_,h(Z,\_))), q(Z,\_)}\] 
where $\mbox{\tt \_}$\
denotes an anonymous variable. Then, $nodes(msg(Q_1,Q_2)) = 6$. Since,
$\vars(Q_1)\cap \vars(Q_2) = \{\mbox{\tt Z}\}$ and 
$\mbox{\tt Z}$ appears twice in the msg, we have $\delta = 2$. Indeed,
$c(Q_1,Q_2) = 8 = 6 + 2 = nodes(msg(Q_1,Q_2)) + \delta$.  
\end{example}

Let us now extend the above definition in such a way that: 1) it deals with goals that do not have an equal number of atoms, 2) it takes commutativity of the $\wedge$ operator into account (that is, we want to consider commonality modulo atom reordering), and 3) it abstracts from concrete variable names in the goals while retaining the sharing information. 
The resulting measure will reflect, for two arbitrary goals, the maximal amount of structure that can be preserved by generalizing a suitable reordering and renaming of both goals.

When comparing two arbitrary goals, we will focus on those subgoals that are {\em similarly structured}, that is, subgoals that correspond to sets of atoms that basically represent calls to the same predicates. More formally:

\begin{definition}
Let $Q_1$ and $Q_2$ be two goals. We say that $(Q'_1,Q'_2)$ is a pair of {\em similarly structured subgoals} of $Q_1$ and $Q_2$ iff $Q'_1$ is a subgoal of $Q_1$, $Q'_2$ is a subgoal of $Q_2$, and $\Pi(Q'_1) = \Pi(Q'_2)$ where $\Pi(Q)$ denotes the multiset of predicate symbols occurring in a goal $Q$. 
\end{definition}
Note that by definition, two similarly structured subgoals always comprise an equal number of atoms as well as an equal number of calls to a particular predicate. Furthermore, we say that a pair $(Q'_1,Q'_2)$ of similarly structured subgoals of $Q_1$ and $Q_2$ is {\em maximal} iff there does not exist another pair of similarly structured subgoals $(Q''_1,Q''_2)$ of $Q_1$ and $Q_2$ such that $Q'_1\subset Q''_1$ and $Q'_2\subset Q''_2$, where $\subset$ denotes the subgoal relation.
\begin{example}\label{ex:one}
Consider the goals $Q_1 = \mbox{\tt p(a,f(A)), s(A), q(A,B)}$ and
$Q_2 =  \mbox{\tt q(Y,Z), p(f(X),Y), r(Z,S)}$.
The pair $(Q'_1,Q'_2)$ with $Q'_1=\mbox{\tt p(a,f(A)), q(A,B)}$ and 
$Q'_2= \mbox{\tt q(Y,Z), p(f(X),Y)}$ is a maximal pair of similarly 
structured subgoals of $Q_1$ and $Q_2$.
\end{example}
Note that for any given pair of goals, there always exists at least one maximal pair of similarly structured subgoals. If the goals do not contain a call to the same predicate, the (unique) maximal pair of similarly structured goals is $(\Box,\Box)$ with $\Box$ denoting the empty goal. Moreover, since 
goals are considered as multisets  a maximal pair of similarly 
structured subgoals is always unique.
Observe that atoms {\em not} being part of the maximal similarly structured subgoals can never be part of a generalization of both goals. Hence our interest in
maximal similarly structured subgoals.  

When comparing similarly structured subgoals, we would like to abstract from actual variable names, while retaining sharing information between goals. In other words, we would like our measure to return a higher value in the case where one of the goals being compared is a renaming of the other (i.e. it presents the same dataflow). Take for example the goals
\[\begin{array}{lll}
G   & = & \mbox{\tt p(X,Y), q(Y,Z)}\\
G'  & = & \mbox{\tt p(A,B), q(B,C)}\\
G'' & = & \mbox{\tt p(A,B), q(C,D)}\\
\end{array}\]
Since $G'$ is a renaming of $G$, their msg would be a goal identical to either of them (up to renaming). By contrast, the msg of $G$ and $G''$ is identical (up to renaming) to only $G''$. To reflect this observation when measuring the similarity of two goals $Q_1$ and $Q_2$ in our measure, we define a set of variable renamings between the two goals as follows. For convenience, we assume that $\vars(Q_1)\cap \vars(Q_2)=\emptyset$ and that $\# \vars(Q_1) \le \# \vars(Q_2)$, where $\# S$ denotes the cardinality of the set $S$. 

\begin{definition}
Let $(Q_1,Q_2)$ be a pair of similarly structured (sub)goals with $\# \vars(Q_1) \le \# \vars(Q_2)$. We define $R(Q_1,Q_2)$ as the (finite) set of injective mappings from $\vars(Q_1)$ onto $\vars(Q_2)$.
\end{definition}

\begin{example}
Let us take $Q'_1$ and $Q'_2$ as in Example~\ref{ex:one} above. Then we have $\vars(Q'_1) = \{A,B\}$, $\vars(Q'_2)=\{X,Y,Z\}$ and consequently the set $R(Q_1,Q_2)$ comprises the following injective mappings:
\[ R(Q'_1,Q'_2) = \left\{\begin{array}{lll}
A\rightarrow X, B\rightarrow Y & \hspace*{1cm}&
A\rightarrow X, B\rightarrow Z\\
A\rightarrow Y, B\rightarrow X & &
A\rightarrow Y, B\rightarrow Z\\
A\rightarrow Z, B\rightarrow X & & 
A\rightarrow Z, B\rightarrow Y\\
\end{array}\right\}\]
\end{example}


We can now define the commonality between a pair of similarly structured subgoals $Q_1$ and $Q_2$
as the maximal strict commonality one could obtain by changing the order of the atoms\footnote{We call {\em permutation of a goal $G$} any goal obtained by reordering the atoms of $G$.} and renaming:
\begin{definition}\label{def:comonality}
Let $Q_1$ and $Q_2$ be two similarly structured (sub)goals such that $\# \vars(Q_1)\le \# \vars(Q_2)$. The {\em commonality} between $Q_1$ and $Q_2$, which we denote by $C(Q_1,Q_2)$, is defined as:
\[C(Q_1,Q_2) =  max \{c(Q_1\rho,Q'_2)\:|\: Q'_2\mbox{ is a permutation of }Q_2\mbox{ and }\rho\in R(Q_1,Q_2)\}.\]
Similarly, if $Q_1$ and $Q_2$ are such that $\# \vars(Q_1) >  \# \vars(Q_2)$,
then $C(Q_1,Q_2)$ is defined as $C(Q_2,Q_1)$.   
\end{definition}
\alexander{We kunnen nog steeds in problemen geraken met deze definitie als een aantal variableen in $Q_1$ 
gelijk is aan een aantal variabelen in $Q_2$. In dit geval is het niet duidelijk dat $C(Q_1,Q_2)$ dezelfde waarde
zal geven als $C(Q_2,Q_1)$. Commonality moet---mijns inzichts---intuitief symmetrisch zijn.}
\wim{Ik snap je opmerking, maar denk je niet dat commonality effectief symmetrisch is? Of je de ene goal permuteert of de andere zou niets mogen uitmaken en voor die renamings zie ik ook niet direct het probleem... Hoe dan ook, denk je niet dat we dit voor deze versie zo kunnen laten?}

We are now ready to define the {\em similarity} between two arbitrary goals, which we define as the
commonality between the goals' maximal pair of similarly structured subgoals:
\begin{definition}\label{def:similarity}
Let $Q_1$ and $Q_2$ be two goals such that $\# \vars(Q_1)\le \# \vars(Q_2)$. The {\em similarity} between $Q_1$ and $Q_2$, denoted $\sigma(Q_1,Q_2)$, is defined as:
$\sigma(Q_1,Q_2) = C(Q'_1,Q'_2)$, where $(Q'_1,Q'_2)$ is the maximal pair
of similarly structured subgoals of $Q_1$ and $Q_2$. If $Q_1$ and $Q_2$ are 
such that $\# \vars(Q_1) >  \# \vars(Q_2)$, then $\sigma(Q_1,Q_2)$ is defined 
as $\sigma(Q_2,Q_1)$.
\end{definition}

\begin{example}
Let $Q_1,Q_2$ and $Q'_1,Q'_2$ as in 
Example~\ref{ex:one}. For the similarity between $Q_1$ and $Q_2$, we have:
\[\begin{array}{lll}
\sigma(Q_1,Q_2) & = & C(Q'_1,Q'_2)\\
& = & \max \{c(Q'_1\rho,Q''_2)\:|\: Q''_2\mbox{ is a permutation of }Q'_2\mbox{ and }\rho\in R(Q'_1,Q'_2)\}.\\
\end{array}\]
One can easily see that the maximal value is $5$ and it is obtained for 
$\rho = \{A\rightarrow Y, B\rightarrow Z\}$ and $Q''_2 =
\mbox{\tt p(f(X),Y), q(Y,Z)}$. 
\end{example}

From the above example, it can easily be seen that our notion of {\em similarity} between arbitrary goals reflects indeed the maximal amount of structure that one could preserve by taking the most specific generalization of the goals' maximal pair of similarly structured subgoals after renaming and reordering the atoms. 

\begin{corollary}
\label{cor}
Let $Q_1$ and $Q_2$ be arbitrary goals and let $(Q'_1,Q'_2)$ their maximal pair of similarly structured subgoals. Let $Q''_2$ be a permutation of $Q'_2$ as in Definition~\ref{def:similarity} such that $\sigma(Q_1,Q_2) = c(Q'_1\rho,Q''_2)$. Then,
\[ \sigma(Q_1,Q_2) = c(Q'_1\rho,Q''_2) =nodes(msg(Q'_1\rho,Q''_2)) + \delta\]
where $\delta$ denotes the number of identical variables that occur at identical positions in $Q'_1\rho$ and $Q''_2$.
\end{corollary}

In particular, if goals $Q_1$ and $Q_2$ are identical modulo atom reordering and renaming, then $\sigma(Q_1,Q_2)$ equals the {\em total} number of nodes (including the variables) in the term representation of either goal. 

\subsection{Comparing predicate definitions}

We will now extend the notion of similarity from individual goals to complete predicate definitions. When comparing predicate definitions, we will abstract from the order of the arguments in each definition. To that extent, we define the notion of an argument permutation as follows:
\begin{definition}
Given two $n$-ary predicates $p/n$ and $q/n$. An {\em argument permutation} between $p$ and $q$ is a bijective mapping $\{1,\ldots,n\}\mapsto\{1,\ldots,n\}$.
\end{definition}
Note that an argument permutation only exists between predicates having the same arity.
In order to consider two predicate definitions as being similar, we impose the condition that both definitions have the same {\em recursive structure}. By this, we mean that there exists a one-to-one mapping between the clauses in both definitions such that 1) corresponding clauses have the same number of recursive calls and 2) the corresponding recursive calls are identical up to a renaming of the variables, a renaming of the recursive calls and a permutation of the argument positions. Since predicates can be mutually recursive, let us first formally state the notion of a clause mapping between two strongly connected components.
\begin{definition}
Let $[p/n]$ and $[p'/n]$ be two strongly connected components. A {\em clause mapping} between $[p/n]$ and $[p'/n]$ is a bijective mapping $\varphi$ with $\dom(\varphi) = \Clauses([p/n])$ and $range(\varphi)=\Clauses([p'/n])$ such that 
for any clauses $c_1, c_2\in \Clauses([p/n])$, we have
$\pred(\head(c_1)) = \pred(\head(c_2))\Leftrightarrow \pred(\head(\varphi(c_1))) = 
\pred(\head(\varphi(c_2)))$.
\end{definition}
A clause mapping establishes a 1-1 correspondence between the clauses of two strongly connected components. Note that such a clause mapping implicitly defines a bijective mapping between the {\em predicates} of the components. Slightly abusing notation, if $\varphi$ is a clause mapping between $[p/n]$ and $[p'/n]$ and $q/m\in [p/n]$, we will use $\varphi(q)$ to denote the {\em predicate} in  $[p'/n]$ whose definition corresponds (by $\varphi$) to the definition of $q$.
\begin{example}\label{ex:clausemapping}
Consider the predicates {\tt append/3} and {\tt concat/3} from the introduction. The mapping $\varphi$ mapping the $i$'th clause of {\tt append/3} onto the $i$'th clause of {\tt concat/3} (for $i=1,2$) is a clause mapping between [{\tt append/3}] and [{\tt concat/3}].
\end{example}

Given a clause mapping between two strongly connected components, we can formally state the conditions under which both components are considered to have the same recursive structure.
\begin{definition}\label{def:same_recursive_structure}
Let $[p]$ and $[p']$ be two strongly connected components and $\varphi$ a clause mapping between $[p]$ and $[p']$. We say that $[p]$ and $[p']$ {\em have the same recursive structure w.r.t. $\varphi$} if and only if the following holds: 
1) for any predicate $q\in [p]$ there exists an argument permutation $\pi_q$ between $q$ and $\varphi(q)$ and 2) for any clause $c\in\Clauses([p])$ of the form
\[A_0\leftarrow Q_1,A_1,\ldots,Q_k,A_k,Q_{k+1}\]
the corresponding clause $\varphi(c)\in [p']$ is of the form
\[A'_0\leftarrow Q'_1,A'_1,\ldots,Q'_k,A'_k,Q'_{k+1}\]
and there exists a variable renaming $\rho$ of $c$ such that for every $A_i$ (with $0\le i\le k$) we have that if $A_i = q(t_1,\ldots,t_m)$ for some predicate $q/m$, then $A'_i = \varphi(q)(t_{\pi_q(1)},\ldots,t_{\pi_q(m)})\rho$.
\end{definition}

The above definition implies that two {\em predicates} have the same recursive structure if there exists a clause mapping between them such that the corresponding clauses contain the same number of recursive calls and there exists an argument permutation that renders the corresponding calls identical (modulo a variable renaming). The same must hold for the heads of the corresponding clauses. When considering strongly connected components rather than individual predicates, the same must hold for each pair of corresponding predicates.

\begin{example}\label{ex:same_recursive_structure}
The {\tt append/3} and {\tt concat/3} predicates from the introduction have the same recursive structure. Indeed, take the clause mapping $\varphi$ from Example~\ref{ex:clausemapping} and take for the argument permutation between {\tt append/3} and {\tt concat/3} the mapping $\pi = \{(1,2), (2,3), (3,1)\}$ and for the renaming $\rho = \{X/E, Xs/Es\}$.
\end{example}

Note that Definition~\ref{def:same_recursive_structure} is not restricted to recursive predicates. 
Two {\em non-recursive} predicates are characterized as having the same recursive structure if there exists an argument permutation between both predicates that makes the heads of the corresponding clauses identical modulo renaming.
Also note that, in principle at least, there might exist several clause mappings between two predicates (or SCCs) under which the predicates (or SCCs) have the same recursive structure.

We are now ready to define the similarity between two strongly connected components. As said before, we only consider strongly connected components that have the same recursive structure.

\begin{definition}
Let $[p]$ and $[p']$ be two strongly connected components that have the same recursive structure w.r.t. a clause mapping $\varphi$. The {\em similarity} between $[p]$ and $[p']$ w.r.t. $\varphi$, denoted by $\sigma([p],[p'],\varphi)$, is defined as 
\[\sigma([p],[p'],\varphi) = \sum_{c\in\Clauses([p])}\: (1+\mbox{$\sum_{i=1}^{k+1} \sigma(Q_i,Q'_i)$} + \mbox{$\sum_{i=0}^k c(A''_i,A'_i))$}\]
if $c$ and $\varphi(c)$ are clauses of the form 
\[A_0\leftarrow Q_1,A_1,\ldots,Q_k,A_k,Q_{k+1}\]
and
\[A'_0\leftarrow Q'_1,A'_1,\ldots,Q'_k,A'_k,Q'_{k+1}\]
respectively and $A''_i$ is $q'(t_{\pi_q(1)},\ldots,t_{\pi_q(m)})\rho$,
if $A_i=q(t_1,\ldots,t_m)$, $pred(A'_i)=q'$, and $\pi_q$ and $\rho$ refer to the required argument permutation and renaming from Definition~\ref{def:same_recursive_structure}.
\end{definition}

In other words, the similarity between two predicate definitions (or SCCs) is defined as the sum of the similarities between the corresponding clauses; the similarity between a pair of clauses comprises two main parts: 1) sum of the similarities between each pair of corresponding {\em non-recursive} subgoals, and 2) the sum of the commonalities between the heads and the corresponding {\em recursive} subgoals. Note that in order to compute the latter, we need to account for the difference in predicate names and the possible permutation of the arguments (hence the use of $A''_i$). 
Also note that, for each clause, we add 1 to reflect the node represented by the~:-~functor in the clause's term representation. One can show that a statement
similar to Corollary~\ref{cor} holds for clauses and strongly connected
components.

\begin{example}\label{ex:appconcat}
Let us reconsider the {\tt append/3} and {\tt concat/3} predicates from the introduction. Their definitions have the same recursive structure w.r.t. $\varphi$
from Example~\ref{ex:same_recursive_structure} (to see this take $\pi_{\tt app}$ and $\rho$ as in Example~\ref{ex:same_recursive_structure}). None of their clauses contain non-recursive subgoals, hence when computing the similarity between both definitions, we sum, for each clause, the commonalities between the heads and recursive calls. We have
\[\begin{array}{lll}
c(\mbox{\tt concat(L,[],L)},\mbox{\tt concat(L,[],L)}) & = & 4\\
c(\mbox{\tt concat([E|Zs],[E|Es],Y)},\mbox{\tt concat([E|Zs],[E|Es],Y)}) & = & 8\\
c(\mbox{\tt concat(Zs,Es,Y)}, \mbox{\tt concat(Zs,Es,Y)}) & = & 4\\
\end{array}\]
Hence, we obtain $\sigma([\mbox{\tt append}], [\mbox{\tt concat}], \varphi) = (1+4) + (1+8+4) = 18$. 
\end{example}

\begin{example}\label{ex:rev}
Consider the {\tt rev\_all} and {\tt add1\_and\_sqr} predicates from the introduction. One can easily verify that both predicates have the same recursive structure; the required clause mapping, argument permutation and renaming are all the identical mapping. With respect to the similarity between the two definitions, it is clear that the corresponding {\em non-recursive} subgoals have no similarly structured subgoals, hence we have
\[\begin{array}{lll}
c(\mbox{\tt add1\_and\_sqr([],[])},\:\: \mbox{\tt add1\_and\_sqr([],[])}) & = & 3\\
c(\mbox{\tt add1\_and\_sqr([X|Xs],[Y|Ys])},\:\: \mbox{\tt add1\_and\_sqr([X|Xs],[Y|Ys])}) & = & 7\\
c(\mbox{\tt add1\_and\_sqr(Xs,Ys)},\:\: \mbox{\tt add1\_and\_sqr(Xs,Ys)}) & = & 3\\
\sigma(\mbox{\tt reverse(X,Y)},\:\: \mbox{\tt (N is X+1, Y is N*N)}) & = & 0\\ 
\end{array}\] 
Hence, we obtain 
$\sigma([\mbox{\tt rev\_all}], [\mbox{\tt add1\_and\_sqr}], \varphi) = (1+3)+(1+7+3+0) = 15$.
\end{example}


Intuitively, it is clear that the notion of similarity between predicate definitions represents the number of nodes that are common to both term representations of the involved predicates. Our notion is quite liberal in the sense that it allows for: 1) renaming of the involved predicate and variable names, 2) permutation of the arguments, and 3) permutation of the body atoms within each non-recursive subgoal. Moreover, by relating the similarity between two predicate definitions to the total number of nodes that are effectively present in each of the definitions' term representations, we obtain an indication of how {\em close} each definition is to some most specific generalization of both definitions.

\begin{definition}
Let $[p]$ and $[p']$ be two strongly connected components that have the same recursive structure with respect to some clause mapping $\varphi$. The {\em closeness} between $[p]$ and $[p']$, denoted $\gamma([p],[p'])$, is defined as the pair
\[\left(\frac{m}{N_{[p]}}\,,\, \frac{m}{N_{[p']}}\right).\]
where $m=\sigma([p],[p'],\varphi)$ and $N_{[p]}$ (or $N_{[p']}$) represent the total number of nodes in the term representations of the predicates in $[p]$ (or $[p']$).
\end{definition}
\noindent
Note that the closeness as defined by the definition above is a pair of values between 0 and 1. Also note that it is (1,1) in case the predicates under consideration are duplicates.  


\begin{example}\label{ex:mp}
One can easily verify that the (total) number of nodes in both the term representations of the {\tt append} and {\tt concat} definitions is 18. Therefore, from Example~\ref{ex:appconcat} it follows that the closeness between them is $(1,1)$, indicating they are duplicates. The number of nodes in {\tt rev\_all} and {\tt add1\_and\_sqr} is, respectively, 19 and 25. By Example~\ref{ex:rev}, it follows that the closeness between them is $(0,79, 0.6)$. These numbers indicate how {\em close} each of these definitions is to the code structure that is common to both of them, which we could represent by the following definition:
\[\mbox{\begin{SProg}
mp(A,B):- A = [], B = [].\\	
mp(A,B):- A = [X|Xs], B = [Y|Ys], mp(Xs,Ys).\\
\end{SProg}}\]
\end{example}

Generalizing the examples given above, we conjecture that, under certain conditions, the closeness between predicates is a useful indication on how much duplicated code is contained in their definitions.

\subsection{Discussion}

In the preceding sections, we have defined the notions that allow to characterize the similarity between predicate definitions. A necessary condition for predicates to be considered similar is that they have the same recursive structure. Definition~\ref{def:same_recursive_structure} requires that for each pair of corresponding clauses, the corresponding recursive calls (or heads) contain the same terms as arguments (modulo an argument permutation and variable renaming). While this might seem overly restrictive, a possible remedy is to compute similarities on programs in a normal form where each atom is of the form: $p(X_1, \ldots, X_n)$, $X = Y$ or $X = f(X_1, \ldots, X_n)$ (with $X,Y,X_1,\ldots,X_n$ different variables). Let us reconsider the {\tt append} and {\tt concat} definitions, this time in normal form:
\[\mbox{
\begin{SProg}
append(X,Y,Z):- X = [], Z = Y.\\
append(X,Y,Z):- X = [Xe|Xs], Z = [Xe|Zs], append(Xs,Y,Zs).\\
\ \\
concat(A,B,C):- B = [], A = C.\\
concat(A,B,C):- A = [Be|As], B = [Be|Bs], concat(As,Bs,C).\\
\end{SProg}}\]
Note that these definitions still have the same recursive structure. Also note that although the computed similarity values will somewhat change due to the presence of the extra body atoms, the similarities will remain identical for both predicates, and thus the closeness between them will still be $(1,1)$, indicating they are duplicates. Changing the order of the unifications in one of the definitions does not influence the computed numbers as these are independent of the order of the body atoms in the non-recursive subgoals. However note that our definitions only capture permutations of body atoms that are confined within a single non-recursive subgoal. Take for example the definition of {\tt append} from above, where we move the unification {\tt Z = [Xe|Zs]} over the recursive call:
\[\mbox{
\begin{SProg}
append(X,Y,Z):- X = [], Z = Y.\\
append(X,Y,Z):- X = [Xe|Xs], append(Xs,Y,Zs), Z = [Xe|Zs].\\
\end{SProg}}\]
By Definition~\ref{def:same_recursive_structure}, this version of {\tt append} still exhibits the same recursive structure as the {\tt concat} predicate above. Nevertheless, the similarity between the definitions will be significantly lower, since the corresponding non-recursive subgoals contain a different number of unifications. 

We believe that restricting the computation of similarity to corresponding non-recursive subgoals does not impose a real limitation. In fact, moving a computation over a recursive call usually represents a significant change in program (and computation) structure that goes beyond the changes in program structure that we would like our technique to be able to detect. 

The computation of similarities lends itself to a {\em top-down} calculation. Indeed, one can first compute what predicates have the same recursive structure. Next, for each pair of predicates having the same recursive structure one can compute the similarities between each pair of corresponding non-recursive subgoals. Complexity of such an algorithm is quadratic in the number of predicates. 

In the following section, we present a more efficient technique approximating the computation of similarities. The idea is to compute, for each predicate definition in isolation, a so-called {\em fingerprint}. Such a fingerprint captures in a single value those characteristics of a predicate that are significant when searching for duplicated (or common) functionality while it abstracts from those characteristics that are less relevant during the search. The computation of these fingerprints does not require any comparison between the definitions of different predicates. Comparing fingerprints is considerably easier than comparing predicate definitions and we believe that the result provides a useful indication about what predicates are possible candidates for a more thorough comparison~\cite{wim:lopstr2004}.

\section{Fingerprinting logic programs}
In what follows, we will map a predicate to a so-called {\em fingerprint}.  The fingerprint of a predicate is a value that is constructed in such a way that (1) it reflects the recursive structure of the predicate, (2) predicates that are duplicates are mapped onto the same value, and (3) the more predicates are similar, the closer the values of their fingerprints. Clearly, fingerprints can be
seen as abstractions (cf.\ ~\cite{CousotCousot76-1,CousotCousot77-1,Cousot:Cousot}. In what follows, we consider programs in normal form as defined above; that is, every atom in the program is of the form $p(X_1,\ldots,X_n)$, $X=Y$ or $X=f(Y_1,\ldots,Y_n)$.
We proceed in a stepwise fashion and define domains of fingerprints over goals, clauses and predicates. For each category we define an order relation over the introduced domain.

The basic idea behind our fingerprinting technique is to abstract a goal by counting the number of occurrences of each function and predicate symbol. 

\begin{definition}
Let $\mathcal A$ be an alphabet, and $F_{\mathcal A}$, 
$\Pi_{\mathcal A}$, ${\mathcal Q}_{\mathcal A}$ respectively, the 
corresponding sets of function symbols, predicate symbols and normalized 
goals. The goalprint function $\varphi_g$ associates every goal  
$Q\in {\mathcal Q}_{\mathcal A}$ with a total 
function $\varphi(Q): (F\cup \Pi\cup \{=\}) \mapsto \mathbb{N}$,
called the goalprint of $Q$, such that:
\begin{itemize}
\item $\varphi_g(p(Y_1,\ldots,Y_n))(h) = 1$ if $h$ is $p$ and $0$, otherwise;
\item $\varphi_g(X = f(Y_1,\ldots,Y_n))(h) = 1$ if $h$ is $f$ or $h$ is $=$, and $0$, otherwise; 
\item $\varphi_g(Q_1\wedge Q_2)(h) = \varphi_g(Q_1)(h) + \varphi_g(Q_2)(h)$ for all $h\in F\cup \Pi$. 
\end{itemize}
The set of all goalprints over $\mathcal A$, i.e., $\varphi_g({\mathcal Q}_{\mathcal A})$, is denoted $\gpA$. 
\end{definition}

Computing the goalprint associated to a goal is straightforward given that the predicate definitions are in normal form. Observe that by computing a goalprint, we ignore the order of the atoms in the goal and the sharing between them. For a given alphabet, all goalprints range over the same domain. Hence, we define the following (total) order on \gpA: Let $\varphi_1, \varphi_2\in\gp$, we say $\varphi_1\preceq\varphi_2$ if and only if $\forall f\in(\dom(\varphi_1)=\dom(\varphi_2))$ we have that $\varphi_1(f)\le \varphi_2(f)$. 

\begin{example}\label{ex:goalprint}
Consider the goal $X=[A|As], As = [B|Bs], p(A,B,C)$. A goalprint $\varphi$ for this goal would be 
$\varphi=\{([|],2), ((=),2), (p,1)\}$.\footnote{We leave the alphabet implicit and assume that a goalprint associates 0 to every function or predicate symbol that is not explicitly mentioned.}
\end{example}

\noindent
Let us conjecture the following result, relating the greatest lowerbound of two goalprints to the generalization (and thus similarity) of the concerned goals.

\begin{conjecture}\label{conj:glb}
Let $Q_1$ and $Q_2$ be arbitrary goals and let $(Q'_1,Q'_2)$ their maximal pair of similarly structured subgoals. Let $Q''_2$ be a permutation of $Q'_2$ and $\rho$ a renaming as in Definition~\ref{def:comonality} such that $\sigma(Q_1,Q_2) = c(Q'_1\rho,Q''_2)$. Then we have
\[\varphi(Q_1) \sqcap \varphi(Q_2) = \varphi(msg(Q'_1,Q''_2)).\]
\end{conjecture}

This can easily be seen, as the greatest lowerbound $\varphi(Q_1) \sqcap \varphi(Q_2)$ indicates precisely the number of occurrences of each predicate and function symbol shared by $Q_1$ and $Q_2$ (and which are hence part of their most specific generalization). As a special case, note that if $nodes(msg(Q'_1,Q''_2)) = nodes(Q_1) = nodes(Q_2)$, that then $\varphi(Q_1) = \varphi(Q_2)$. In other words, duplicated goals will have identical goalprints. Note that the converse does not necessarily hold: 
\begin{example}
the goals $Q_1:$ {\tt X = f(Y), Y = f(Z)} and $Q_2:$ {\tt A = f(B), C = f(B)}
have identical goalprints, yet $\sigma(Q_1,Q_2)$ does not equals $nodes(Q_1)$ nor $nodes(Q_2)$. Indeed,  $\sigma(Q_1,Q_2) = 7$ whereas $nodes(Q_1)=nodes(Q_2)=9$.
\end{example}


We will now use the notion of goalprint to construct fingerprints of clauses and predicates.
When abstracting a single clause of the form
\[A_0\leftarrow Q_1,A_1,\ldots,Q_k,A_k,Q_{k+1}\]
we keep track of the individual abstractions of the non-recursive subgoals $Q_i$. Therefore, we define the fingerprint of a clause as a {\em sequence} of goalprints: one for every (maximal) non-recursive subgoal of the clause body.

\begin{definition}
Let $\mathcal A$ be an alphabet and $P_{\mathcal A}$ be a program over the alphabet. 
A {\em clauseprint function} $\varphi_c$ maps 
every clause $A_0\leftarrow Q_1,A_1,\ldots,Q_k,A_k,Q_{k+1}$
to a sequence of goalprints $\langle \varphi_g(Q_1),\ldots,\varphi_g(Q_{k+1})\rangle$,
called a {\em clauseprint}. 
The set of all such clauseprints, i.e., $\varphi_c(P_{\mathcal A})$ is denoted by \cpA. That is
$\cpA\subseteq \gpA^*$. 
\end{definition}

Since our primary interest is in comparing clauses having the same recursive structure, we define the following 
partial order on clauseprints. Let $\varphi^*_1,\varphi^*_2\in\cpA$; we define $\varphi^*_1\preceq\varphi^*_2$ if 
and only if $\varphi^*_1=\langle \varphi_1,\ldots,\varphi_n\rangle$ and $\varphi^*_2=\langle \varphi'_1,
\ldots,\varphi'_n\rangle$ for some $n\in\mathbb{N}$ and  $\varphi_i\preceq\varphi'_i$ for all $1\le i\le n$.

\begin{example}\label{ex:clauseprint}
Reconsider the definitions of {\tt append} and {\tt concat} in normal form. The first clause of both predicates can be characterized by the clauseprint $\langle \varphi_1\rangle$ with $\varphi_1 = \{([],1), ((=),2)\}$; the second clauses by $\langle \varphi_{2,1},\varphi_{2,2}\rangle$ with $\varphi_{2,1}=\{([|],2),((=),2)\}$ and $\varphi_{2,2} = \{\}$.
\end{example}

Finally, the fingerprint of a predicate is defined as a function associating a clauseprint to each of the clauses in the predicate's definition.

\begin{definition}
Let $\mathcal A$ be an alphabet, let $P_{\mathcal A}$ be a program over the alphabet and $\Pi_P$ be the set
of predicates in the program. {\em Predicate print function} $\Phi$ 
maps every predicate $p\in \Pi_P$ to a multiset
$\multiset{\varphi_c(c)\mid c\in \Clauses(p)}$, called a {\em predicate print}. The set of all predicate
prints is denoted $\ppA$.
\end{definition}

Observe that we use multisets rather than sets since different clauses of the same predicate can give rise to identical clauseprints. In this case
we would like the clauseprint to appear twice in the predicate print.

\begin{example}\label{ex:fp_duplicates}
The predicate prints of {\tt append} and {\tt concat} from before, denoted $\Phi_{app}$ and $\Phi_{conc}$ are defined as
$$\Phi_{app} = \Phi_{conc} = \multiset{\langle \varphi_1\rangle, \langle \varphi_{2,1},\varphi_{2,2}\rangle},$$
with $\varphi_1, \varphi_{2,1}$ and $\varphi_{2,2}$ as in Example~\ref{ex:clauseprint}.
\end{example}

We define the following partial order on \ppA. Let $\Phi_1,\Phi_2\in\ppA$; we define $\Phi_1\preceq\Phi_2$ if and only if $\Phi_1(c)\preceq\Phi_2(c)$ for all $c\in \Pi_P$. Observe once again that the order relation $\preceq$ is only defined between fingerprints of predicates having the same recursive structure.
As a final example, let us reconsider the predicates {\tt rev\_all} and {\tt add1\_and\_sqr} in normal form:
\begin{example}\label{ex:fp_common}
\[\mbox{\begin{SProg}
rev\_all(A,B):- A = [], B = [].\\
rev\_all(A,B):- A = [X|Xs], B = [Y|Ys], reverse(X,Y), rev\_all(Xs,Ys).\\
\\
add1\_and\_sqr(A,B):- A = [], B = [].\\
add1\_and\_sqr(A,B):- A = [X|Xs], B = [Y|Ys], N is X + 1, Y is N*N, \\
\qin\qin \qin\qin\qin\qin\qin\qin\  add1\_and\_sqr(Xs,Ys).
\end{SProg}}\]
The associated predicate prints are $\Phi_{ra}$ and $\Phi_{aas}$ defined as:
\[\begin{array}{l}
\Phi_{ra} = \multiset{\langle \{([],2),((=),2)\}\rangle, \langle \{([|],2),((=),2),(reverse,1)\}, \{\} \rangle}\\
\Phi_{aas} = \multiset{\langle \{([],2),((=),2)\}\rangle, \langle \{([|],2),((=),2),(is,2),(+,1),(*,1)\}, \{\} \rangle}\\
\end{array}\]
Both predicate prints are comparable and computing their greatest lowerbound, i.e. $\Phi = \Phi_{ra}\sqcap\Phi_{aas}$ gives us the following predicate print
$\Phi = \multiset{\langle \{([],2),((=),2)\}\rangle, \langle \{([|],2),((=),2)\}, \{\} \rangle}$,
which corresponds indeed to the fingerprint of the {\tt mp/2} predicate from Example~\ref{ex:mp}, reflecting the common code structure of the {\tt rev\_all} and {\tt add1\_and\_sqr} predicates.
\end{example}


Similarly, we define an SCC-print function $\Phi$ and an SCC-print for an SCC $[p]$ as a multiset of predicate prints corresponding to all predicates in $[p]$.

\begin{lemma}
Let $[p]$ and $[p']$ be two strongly connected components that have the same recursive structure with respect to some clause mapping $\varphi$. 
If $\gamma([p], [p']) = (1,1)$ then $\Phi([p]) = \Phi([p'])$.
\end{lemma}

\section{Discussion and ongoing work}

Conjecture~\ref{conj:glb} relates the greatest lowerbound of two goalprints to the most specific generalization of the goals (after renaming and atom reordering) and thus to their similarity. An interesting topic of future work is to extend these results to complete predicate definitions. Doing so requires a formal characterization of the most specific generalization of two predicates (as always modulo renaming and atom reordering). As suggested by Examples~\ref{ex:fp_duplicates} and~\ref{ex:fp_common}, we conjecture that the greatest lowerbound of two predicate prints neatly characterizes the similarity between the predicates and thus their common code. Literal code duplication is reduced to a special case since duplicated predicates (having closeness (1,1)) have identical fingerprints.

Other topics of future work include to adapt the techniques proposed in this paper to Prolog. This requires, among others, to constrain the notion of clause mapping (to fix the order in which clauses must be mapped onto each other) and to limit the amount of reordering permitted when computing the similarity between non-recursive subgoals.
Finally, we intend to investigate the relation with fingerprinting techniques used for the detection of plagiarism, like e.g. \cite{schleimer03winnowing}, and to make a prototype implementation of our proposed technique and to evaluate its effectiveness and performance on a testbed of programs.

\bibliographystyle{plain}
\bibliography{submission}
\end{document}